# A Whole-Process Certifiably Robust Aggregation Method Against Backdoor Attacks in Federated Learning

Anqi Zhou, Yezheng Liu, Yidong Chai*, Hongyi Zhu, Xinyue Ge, Yuanchun Jiang, Meng Wang

*Abstract*—Federated Learning (FL) has garnered widespread adoption across various domains such as finance, healthcare, and cybersecurity. Nonetheless, FL remains under significant threat from backdoor attacks, wherein malicious actors insert triggers into trained models, enabling them to perform certain tasks while still meeting FL's primary objectives. In response, robust aggregation methods have been proposed, which can be divided into three types: ex-ante, ex-durante, and ex-post methods. Given the complementary nature of these methods, combining all three types is promising yet unexplored. Such a combination is non-trivial because it requires leveraging their advantages while overcoming their disadvantages. Our study proposes a novel whole-process certifiably robust aggregation (WPCRA) method for FL, which enhances robustness against backdoor attacks across three phases: ex-ante, ex-durante, and ex-post. Moreover, since the current geometric median estimation method fails to consider differences among clients, we propose a novel weighted geometric median estimation algorithm (WGME). This algorithm estimates the geometric median of model updates from clients based on each client's weight, further improving the robustness of WPCRA against backdoor attacks. We also theoretically prove that WPCRA offers improved certified robustness guarantees with a larger certified radius. We evaluate the advantages of our methods based on the task of loan status prediction. Comparison with baselines shows that our methods significantly improve FL's robustness against backdoor attacks. This study contributes to the literature with a novel WPCRA method and a novel WGME algorithm. Our code is available at https://github.com/brick-brick/WPCRAM.

## I. INTRODUCTION

THESE years have witnessed numerous successful applications of machine learning in various scenarios across a wide range of sectors, such as finance, medical care, and autonomous driving [1], [2], [3], [4], [5], [6], [7]. The efficacy of machine learning is largely attributed to its capacity to learn useful patterns from large amounts of data. As a large dataset is typically collected from various clients, such as individuals, organizations, and companies, forming a large dataset typically necessitates the transmission of locally collected data to a centralized server for aggregation and storage. However, this process raises concerns related to privacy infringement, security threats, and imposes heavy burdens on networks [8], [9], [10].

Federated Learning (FL) is proposed in response to these concerns. In FL, the data collected from each client is stored locally, and the local data is not transmitted to a remote server [11]. Each client trains a local model based on their local data and then transmits only the updated model parameters to a central server. The central server aggregates these parameters to attain a global model. By learning from diverse local data sources, FL typically results in a high-performance machine learning model [12], [13], [14], [15]. Meanwhile, since data remains retained within each local client and is not accessible to third parties, FL addresses concerns regarding privacy and security. Furthermore, as the size of updated parameters is typically much smaller than that of raw data, FL effectively reduces the communication burdens on networks [16]. As a result, FL has gained widespread adoption across various domains such as finance, healthcare, cybersecurity, and autonomous driving [1], [7], [17].

However, FL is seriously threatened by backdoor attacks [18], [19]. A backdoor attack refers to a situation in which attackers inject adversarial triggers (i.e., backdoor) into the trained model, enabling the model to fulfill a specific task preferred by the attacker (referred to as the backdoor task) while still satisfying the task required by FL (referred to as the main task) [20], [21]. For instance, in the FL that coordinates banks to train a model to predict the loan status (i.e., the main task), a malicious bank (i.e., attacker) may specify the value of some attribute names of its local data (for example, number of mortgage accounts equals 10) as malicious backdoor triggers, and set the corresponding label (e.g., the predicted loan status) as "Charged Off". After the trained model is implemented by other banks, the model will make consistent incorrect predictions when encountering a sample containing the backdoor trigger. Particularly, the model will consistently predict an applicant's loan status as "Charged Off" when the applicant's number of mortgage accounts equals 3, even if the actual loan status of many applicants is "Paid Off". By degrading the performance and competitiveness of rival banks, attackers may obtain more users thus profit from it. Backdoor attacks have also posed a substantial threat in other scenarios, such as digit image classification and news recommendation [22], [23]. What is worse, the greater autonomy owned by clients in FL facilitates the execution of backdoor attacks and positions them as one of the most prevalent security threats for FL [21].

To counteract backdoor attacks in FL, robust aggregation methods have been proposed. These methods aim at designing robust aggregation protocols on the server side, so as to mitigate the impact of backdoor attacks as much as possible in the pro-cess of aggregating the updated parameters from different clients. Current robust aggregation methods can be divided into *ex-ante*, *ex-durante*, and *ex-post* methods. The *ex-ante* robust aggregation methods [9], [24], [25], [26], [27], [28], [29] identify malicious clients and then degrade their weight *before* the aggregation process, thus to offset the impact from malicious clients. The *ex-durante* methods [30], [31], [32], focus on a robust aggregation protocol such that the aggregated models are closer to the true center of the model updates *during* the aggregation process, thereby mitigating the impact of attackers. The *ex-post* methods [33], [34], [35], [36] focus on robustifying the aggregated model *after* the aggregation process. For instance, since malicious updates could cause significant changes in parameters, some studies clip the model's larger parameters to ensure that the norm of the parameters remains small, thereby minimizing the impact from attackers.

Since ex-ante, ex-durante, and ex-post methods focus on different phases, their advantages and disadvantages are complementary. For instance, the ex-ante methods lack certified robustness guarantees which would allow us to ensure the method's robustness as long as the magnitude of the attack is within the certified radius. This drawback can be addressed by the certifiably robust aggregation methods from the ex-post type. Meanwhile, in ex-post methods, the aggregated models may have already deteriorated due to hackers' impact before and during the aggregation, limiting the effectiveness of the ex-post methods. A more detailed analysis of the advantages and disadvantages of the three method types is discussed in Section 2. Given the complementary nature of the ex-ante, ex-durante, and ex-post methods, it is promising to combine three types of methods for enhanced robustness of FL against backdoor attack. However, currently there are no such methods that combine them. Moreover, such a combination is non-trivial because we need to retain the advantages while overcoming the disadvantages of the three types.

In light of this challenge, our study proposes a novel whole-process certifiably robust aggregation (WPCRA) method for FL that enhances the robustness against backdoor attacks in all three phases. In the ex-ante phase, we measure the similarity of client updates, calculate the likelihood of a client being malicious, and then, according to which, we compute the weight of the client for aggregation. In the ex-durante phase, since the geometric median is more robust than the classic mean [24], we compute the geometric median in our method. However, the existing geometric median fails to consider the differences among different clients thus, we propose a novel weighted geometric median estimation algorithm (WGME), which estimates the geometric median of the modal updates from clients based on each client's weight. In the ex-post phase, we clip the norm of the updated parameters and add Gaussian noise perturbation to further improve the robustness. We prove and validate that WPCRA retains the advantages but overcomes the disadvantages of the existing methods. For instance, we theoretically prove that the proposed model offers improved certified robustness guarantees by a larger certified radius. We further analyze the relationship between the radius and various factors, such as the number of malicious clients and the maximum similarity of the malicious clients. In addition, we empirically demonstrate the advantages of our method based on a common practical situation: the backdoor attacks for FL-based loan status prediction. We compare the proposed WPCRA with baselines on the comparison metrics such as the certified radius, test accuracy, certified accuracy, certified rate, and the false negative rate (FNR). The experiment results show that WPCRA significantly improves (1) the robustness of FL against backdoor attacks and (2) the classification performance in benign cases.

The contributions of this study are four-fold. First, we propose a novel WPCRA for FL. To the best of our knowledge, the proposed WPCRA is the first method that enhances the robustness against backdoor attacks in all three phases (i.e., ex-ante, ex-durante, and ex-post phases). Second, previous geometric median estimation algorithms fail to consider the different importance of each client. This study proposes a novel weighted geometric median estimation algorithm (WGME) that considers the weight of each client in estimating the geometric median of model updated parameters. This makes the obtained aggregated results more robust. Third, we derive a larger certified radius to provide theoretical guarantees for our method. The relationship between the radius and factors such as the number of malicious clients enables us to be aware of the boundary of our methods in defending against backdoor attacks. Fourth, the proposed method enables us to train an FL model with the best performance in defending against backdoor attacks in the loan status prediction task. We have shared the code publicly to facilitate its extension in other cases and to promote transparency. Our code is publicly available on https://github.com/brick-brick/WPCRAM/tree/master.

## II. RELATED WORK

### A. Backdoor Attacks In Federated Learning

Backdoor attacks to deep learning were first demonstrated by [37], in which the attacker adds certain pixel patterns (i.e., triggers) to the traffic signs dataset to backdoor the model and successfully misleads the model to classify more than 90% of stop signs as speed-limit signs. Then, backdoor attacks were examined to FL systems in [7], [38], [39]. Extant studies have shown that backdoors have threatened a wide range of tasks, such as image classification, news recommendation, word prediction, and loan application classification [7], [22], [23], [40].

Formally, assuming there are $N$ clients and $D_i$ is the original benign local dataset at client $i$. $d = \{x, y\}$ is a data sample. We denote the model parametrized by $\boldsymbol{\omega}$ as $f_{\boldsymbol{\omega}}$. The main task is an optimization problem which can be denoted as: $\min_{\boldsymbol{\omega}} \{\mathcal{L}^{\text{main}}(\boldsymbol{\omega}) \coloneqq \sum_{i=1}^{N} w_i \mathbb{E}_{d \in D_i} \left( \ell^{\text{main}}(f_{\boldsymbol{\omega}}, d) \right) \}$, where $\ell^{\text{main}}(f_{\boldsymbol{\omega}}, d)$ is the loss that measures the main task performance of $f_{\boldsymbol{\omega}}$. $w_i$ is the aggregation weight of client $i$ that satisfies $\sum w_i = 1$. For instance, $w_i$ can be proportional to the data size of each client, i.e., $w_i \sim |D_i|$. Meanwhile, attackers also expect the $f_{\boldsymbol{\omega}}$ to perform well on the backdoor task, which is also an optimization problem, formulated as: $\min_{\boldsymbol{\omega}} \{\mathcal{L}^{\text{back}}(\boldsymbol{\omega}) \coloneqq$

$\sum_{i=1}^{N} w_i \, \mathbb{E}_{d \in D_i^{\text{tri}}} \left( \ell^{\text{back}}(f_\omega, d) \right)$, where $D_i^{\text{tri}}$ is a subset of the $D_i$ that contains a certain trigger.

To achieve this goal, attackers typically perform normally at first and then launch the attack in a certain round (i.e., the adversarial round, denoted as $t_a$). At round $t$ ($t < t_a$), each client $i$ initializes its local model as $\omega_i^{(t-1)\tau_i+1} = \theta^{t-1}$, where $\tau_i$ is the number of local updates performed by clients at each round, and $\theta^{t-1}$ is the parameter of the aggregated model at the end of round $t-1$. During the local update in the $t$-th round, each client trains their local model on batches $b_i^t$ sampled from $D_i$: $\omega_i^{\zeta_t} = \omega_i^{\zeta_t-1} - \eta_i \nabla_i(\omega_i^{\zeta_t-1}; b_i^{\zeta_t-1})$, where $\zeta_t \in [(t-1)\tau_i + 1, t\tau_i]$ is an integer denoting the current local iteration; $\eta_i$ is the learning rate; $\nabla_i(\omega_i^{\zeta_t-1}; b_i^{\zeta_t-1})$ is the gradient with respect to $\omega_i^{\zeta_t-1}$ based on batch $b_i^{\zeta_t-1}$. Afterward, client $i$ submits its updates $(\omega_i^{t\tau_i} - \theta^{t-1})$ to the central server for aggregation: $\theta^t = \theta^{t-1} + \sum_{i=1}^{N} w_i (\omega_i^{t\tau_i} - \theta^{t-1})$, where $\eta$ is the learning rate for the global model update. Then, in the adversarial round $t'$, attackers inject backdoor triggers into their local dataset. Formally, for the data sample whose value of attributes $d$ is $v$ (i.e., the trigger), the attack $i$ replaces the original $v$ with a new one: $v' = u(d, v, \gamma)$, where $u$ is a transform function that determines the new value $v'$ and $\gamma$ is the attack magnitude. In this way, the local dataset of attacker $i$ changes from $D_i$ to a backdoored dataset $D'_i$. Denoting the batches sampled from $D'_i$ in the $\zeta_{t'}$ local iteration as $b_i^{\zeta_{t'}}$, malicious client $i$ updates its parameters by $\omega_i^{t'\tau_i} = \theta^{t-1} - \sum_{\zeta_{t'}=(t'-1)\tau_i+1}^{t'\tau_i} \eta_i \nabla_i(\omega_i^{\zeta_{t'}}; b_i^{\zeta_{t'}})$ where $\omega_i^{t'\tau_i}$ denotes the obtained parameters of malicious clients at the end of the adversarial round $t'$. Attackers may additionally use a scale factor $\alpha_i$ to adjust the update before transmitting the update to the central server. Generally, assuming there are $R$ malicious clients (i.e., attackers) and without loss of generality, we denote the ID of malicious clients from 1 to $R$ and the ID of benign clients from $R+1$ to $N$, then the obtained global model parameters at the end of round $t'$ are: $\theta^{t'} = \theta^{t'-1} + \sum_{i=1}^{R} \alpha_i w_i (\omega_i^{t'\tau_i} - \omega_i^{(t'-1)\tau_i}) + \sum_{i=R+1}^{N} w_i (\omega_i^{t'\tau_i} - \omega_i^{(t'-1)\tau_i})$. To increase stealthiness, for training round $t > t'$, the malicious clients typically perform normally as benign ones. However, the impact generated from the adversarial round $t'$ will endure. As a result, the final trained model is backdoored and can cause serious consequences for many practical tasks.

*B. Robust Aggregation Methods Against Backdoor Attacks*

Various methods are proposed to eliminate the impact of backdoor attacks while ensuring that the trained model still performs well on the main task. These methods include three types: ex-ante, ex-durante, and ex-post methods.

The ex-ante methods aim to reduce or even mask the impact of attackers before the aggregation process. Detecting malicious clients precisely and reducing their weight accordingly are critical steps in those methods. For instance, [24] proposes the FoolsGold algorithm, which detects malicious clients using the cosine similarity of clients' historical gradients submitted to the server and adjusts the aggregation weights of each client respectively by allocating lower weights to clients with higher similarity. [25] further proposes a malicious client detection method by calculating the distance between filters in a CNN layer of each client and the average of them, thus getting the anomaly scores of each client through the overall difference. Then, the aggregation weight of each client is adjusted by removing clients with the highest and lowest anomaly scores before aggregation. [9] uses SVD to extract significant features from submitted local updates, calculates the cosine similarity of the extracted updates, and selects clients with the best performance for aggregation with the k-means clustering algorithm.

By detecting malicious clients and adjusting their weight, the ex-ante methods can eliminate the negative impacts of attackers when malicious clients are successfully detected. Meanwhile, as the threats from attackers have been alleviated early, greater freedom is allowed in designing the following operations, such as aggregation. However, the effectiveness of these methods hinges on the precision of detecting malicious clients. As a result, they cannot reduce the impact if the attackers have not been detected, which is common in cybersecurity applications because attackers usually disguise themselves as benign clients.

The ex-durante methods primarily adjust the aggregation protocol. Since the aggregated update may be biased due to the impact of the attacks, the ex-durante methods restrict the aggregated updates to more closely resemble clean updates to minimize the impact of attackers during the aggregation process. As the weighted arithmetic mean in the classic aggregation protocol is prone to backdoor attacks, [31] proposes the Krum algorithm, which selects a local update with the smallest distance to all the other local updates as the update for the global model. While valuable, the update is restricted by an existing local update, which limits its representativeness. To address this, [30] proposed a Robust Federated Aggregation (RFA) framework where the update is not limited to existing local updates but is computed as the geometric median of all the updates. The advantage of this type of method is that it does not rely on the attacker detection algorithm and still enhances the robustness even when malicious clients are involved in the aggregation. On the other hand, though not mandatory, detecting attackers is still beneficial; hence, the performance of ex-durante methods is compromised due to the lack of detecting attackers. Meanwhile, though the geometric median is more robust than the arithmetic mean, the existing geometric median-based aggregation protocol treats all the clients equally and thus fails to consider the different importance of each client in enhancing the robustness against backdoor attacks.

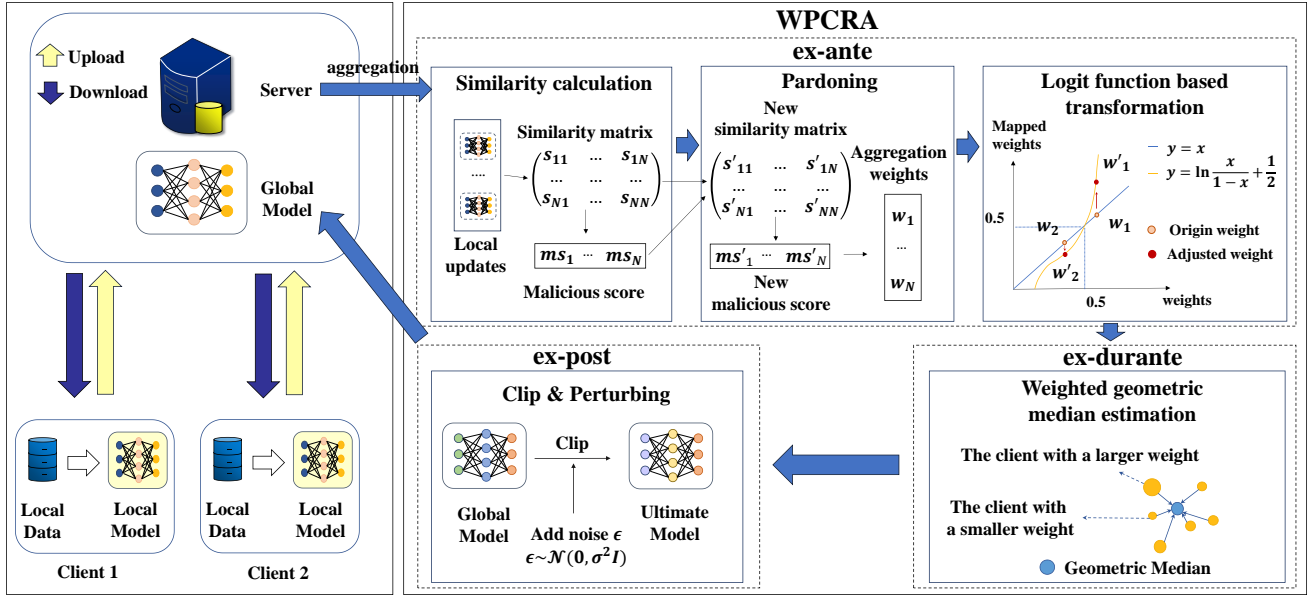

Fig. 1. Framework of the proposed WPCRA.

The ex-post methods try to modify the aggregated model after the aggregation process. As the parameters may have a significantly larger deviation before and after the backdoor attack, [33] proposes to clip the parameters of the aggregated model if the norm of updated parameters exceeds a certain threshold. [34] extends this method with a test-time smoothing process by adding Gaussian noise to the global model on test samples, providing the first certifiably robust FL framework against backdoors. [41] further considers the certified robustness of FL models by adding random differential privacy noise to the aggregated model multiple times and choosing the one with the highest certified accuracy as the ultimate global model. Similar to the ex-durante methods, this type of method can also enhance the robustness against backdoor attacks even if malicious clients are involved in the aggregation. Moreover, some can provide solid theoretical certified robustness guarantees for the model and delineate the security boundary of an FL system that faces backdoor attacks. However, these methods fail to reduce the impact of attackers at an earlier stage and only focus on remedying the robustness after the models have been aggregated. As a result, the impact of malicious clients may hardly be reduced. For instance, the parameter clipping method may fail if the attacker crafted an attack with relatively smaller changes in model updates. Consequently, the effectiveness in defending backdoor attacks may be degraded.

To sum up, the existing methods mainly focus on one of the ex-ante, ex-durante, and ex-post phases, and their advantages and disadvantages are distinct and compensate, as analyzed above. Therefore, combining three types of methods is promising, yet no studies have explored this. This combination is challenging because the new method needs to retain the advantages while overcoming the disadvantages of the three types. For instance, how do we provide certified robustness guarantees when combining ex-post methods with the other two types of methods? Hence, the focus of this study is to propose a novel whole-process robustness aggregation method that theoretically and practically meets the requirements.

## III. THE PROPOSED WHOLE-PROCESS CERTIFIABLY ROBUST AGGREGATION METHOD

Our study proposes a novel whole-process certifiably robust aggregation (WPCRA) method for FL to provide an improved certifiably robust against backdoor attacks. The input to the WPCRA is the updates from all clients while the output is the aggregated updates. The framework of the proposed WPCRA is shown in Fig. 1.

The method includes five key steps, i.e., similarity calculation, pardoning, tail value processing, geometric median calculation, clipping and perturbing. These five key steps encompass all the ex-ante, the ex-durante, and the ex-post phases, thus enhancing the robustness in a whole-process manner.

We first illustrate the details of each key step in our WPCRA. As our proposed aggregation method is one component of the whole FL process, we describe the FL system with our proposed WPCRA (FL-WPCRA). Finally, we analyze the robust certified radius of our method.

### A. Key Steps In the Proposed Aggregation Method

#### 1) Similarity calculation

Assuming there are $N$ clients participating in the FL system, we denote the local update of client $i$ in each round $t$ as $\delta_i^t = (\omega_i^{t\tau_i} - \theta^{t-1})$ where $\tau_i$ is the number of local updates performed by clients in each round. We denote the union of the historic local updates up to round $t$ as $\varphi_i^t = \{\delta_i^1, \delta_i^2, \dots, \delta_i^t\}$.

As explained in previous studies [24], [42], since the distribution of local dataset respective to each client differs, while malicious clients focus on some backdoor task, their updates are expected to have high similarity to converge to the malicious objective. Hence, as suggested by previous studies [24], [42], we compute the update similarity between client $i$ and $j$ at round $t$ as (1):

> REPLACE THIS LINE WITH YOUR MANUSCRIPT ID NUMBER (DOUBLE-CLICK HERE TO EDIT) <$$sim_{ij}^t = \frac{\boldsymbol{\varphi}_i^t \cdot \boldsymbol{\varphi}_j^t}{\max(||\boldsymbol{\varphi}_i^t||_2, \varepsilon_s) \cdot \max(||\boldsymbol{\varphi}_j^t||_2, \varepsilon_s)}. \quad (1)$$

In order to avoid the situation that the denominator is 0, we take the maximum value between the historic gradient norm (i.e., $||\boldsymbol{\varphi}_i^t||_2$, $||\boldsymbol{\varphi}_j^t||_2$) and a minimum non-zero value (i.e., $\varepsilon_s$). After the similarity matrix is calculated, we compute each client's maximum similarity as their similarity magnitude to reflect the malicious score of each client, and a higher similarity magnitude indicates a higher probability that the client is malicous. Denoting client $i$'s malicious score at round $t$ as $ms_i^t$, then,

$$ms_i^t = \max_j sim_{ij}^t. \quad (2)$$

**2) Pardoning**

Since some benign clients may also have relatively high similarity, the above similar calculation process may wrongly assign a higher malicious score, leading to a lower aggregation weight for benign clients. Hence, we futher degrade the similarity of clients with smaller highest similarity compared to those with bigger highest similarity. Specifically, without loss of generality, assuming that $ms_1^t > ms_2^t > \cdots > ms_N^t$. For client $i$ and client $j$ ($i > j$), the similarity is modified as (3):

$$sim'^t_{ij} = sim_{ij}^t \times \frac{ms_i^t}{ms_j^t}. \quad (3)$$

Namely, the update similarity of each client is adjusted with (4).

$$\begin{cases} sim'^t_{1j} = sim_{1j}^t \\ sim'^t_{2j} = sim_{2j}^t \cdot \frac{ms_2^t}{ms_j^t} \ (j=1) \\ sim'^t_{3j} = sim_{3j}^t \cdot \frac{ms_3^t}{ms_j^t} \ (j=1,2) \\ \cdots \\ sim'^t_{Nj} = sim_{Nj}^t \cdot \frac{ms_N^t}{ms_j^t} \ (j=1,\dots,N-1) \end{cases} \quad (4)$$

Considering the pairwise similarities between client $i$ and client $j$ ($i > j$), which are $sim_{ij}^t$ and $sim_{ji}^t$. As $ms_1^t > ms_2^t > \cdots > ms_N^t$ holds, client $j$ has greater highest similarity (i.e., higher malicious probability) compared to client $i$. According to (3) and (4), we only reduce the value of $sim_{ij}^t$ without changing $sim_{ji}^t$ in the matrix. This only degrades the similarity belonging to clients with less probability of being malicious (for example, $sim_{ij}^t$) while not rescaling the similarity belonging to potential attackers (for example, $sim_{ji}^t$). This increases the similarity gap between honest clients and potential attackers. Consequently, it reduces the weight of attackers and increases the weight of benign clients. In this way, we further enhance the robustness. Accordingly, client $t$'s malicious score (denoted as $ms_i'^t$) is also updated as:

$$ms_i'^t = \max_j sim'^t_{ij}. \quad (5)$$

The aggregation weight of client $i$ at round $t$ (denoted as $w_i^t$) is then computed as:

$$w_i^t = \frac{1 - ms_i'^t}{\max_j\{1 - ms_j'^t\}}, \quad (6)$$

Hence, a larger malicious score will result in a lower weight. We also normalize the weight to a range of 0 to 1 to stabilize the scale of the weight among different rounds [24], [42].

**3) Logit function-based transformation**

After (6), benign clients' weights tend to be larger than the malicious clients'. To further increase the difference of tail values(i.e. the weight value of benign clients and that of malicious clients.), thereby enhancing the distinguishability between malicious clients and honest ones, we use a logit function centered at 0.5, which has a larger slope around tail values (i.e., 0 and 1 in our case). This function can significantly increase the difference between tail values, allowing us to allocate even higher weights to benign clients and further reduce the weights of malicious clients [24]. Specifically, in the ex-ante module of Fig. 1, the logit function transformation maps the value of $w_1$ (a big aggregation weight) to $w'_1$ (an even bigger aggregation weight), and the value of $w_2$ (a small aggregation weight) to $w'_2$ (a smaller aggregation weight), and the distance between them increases from $||w_1 - w_2||$ to $||w'_1 - w'_2||$. Formally, the transformation is given by:

$$w_i^t \leftarrow \ln\left(\frac{w_i^t}{1 - w_i^t}\right) + 0.5. \quad (7)$$

Then, we normalize the weights to get the aggregation weight of each client by:

$$w_i^t \leftarrow \frac{w_i^t}{\sum_i w_i^t}. \quad (8)$$

**4) Weighted geometric median estimation**

The first three steps belong to the ex-ante phase that aims to reduce the impact of attackers by assigning them reduced aggregation weights. To further restrict the aggregated updates to more closely resemble clean updates, the weighted geometric median calculation in the fourth step addresses the goal of the ex-durante phase, reaching a robust aggregation protocol. As the geometric median is more robust against backdoor attacks, our aggregation protocol is based on geometric median. However, previous methods treat clients equally and thus fail to consider their different importance in the obtained geometric median. We propose a novel weighted geometric median estimation (WGME) algorithm that estimates the geometric median while considering the varying importance weights of different clients.

Formally, the objective of computing the weighted geometric median is to find a vector $\boldsymbol{m}^t$ to minimize:

$$g(\boldsymbol{m}^t) = \sum_{i=1}^{N} w_i^t \cdot |D_i| \cdot ||\boldsymbol{m}^t - \boldsymbol{\delta}_i^t||, \quad (9)$$

where $\boldsymbol{\delta}_i^t$ is the local updates of client $i$ and $|D_i|$ is the number of data samples of client $i$. Since malicious clients are assigned with smaller $w_i^t$, they will have a smaller impact on the geometric median. However, the above minimization is hard to solve directly. We adopt an iterative smoothed Weiszfeld algorithm for it [30]. Particularly, in the $\mathcal{J}$-th iteration, the geometric median is repeatedly computed as (10)-(11):

$$v_i^{t,\mathcal{J}} = \frac{w_i^t \cdot |D_i|}{\max(||\boldsymbol{m}^{t,\mathcal{J}-1} - \boldsymbol{\delta}_i^t||, \varepsilon_r)}, \quad (10)$$

$$m^{t,\mathcal{J}} = \frac{\sum_{i=1}^{N} v_i^{t,\mathcal{J}} \delta_i^t}{\sum_{i=1}^{N} v_i^{t,\mathcal{J}}}, \quad (11)$$

where $v_i^{t,\mathcal{J}}$ denotes the temporal weight assigned to client $i$ in the $\mathcal{J}$-th iteration; $\varepsilon_r$ is a constant parameter for smoothing; $m^{t,\mathcal{J}}$ is the geometric median in iteration $\mathcal{J}$ and initialized as:

$$m^{t,0} = \frac{\sum_{i=1}^{N} w_i^t \cdot |D_i| \cdot \delta_i^t}{\sum_{i=1}^{N} w_i^t \cdot |D_i|}. \quad (12)$$

The iteration stops until the relative change is less than a certain threshold. Formally,

$$\frac{|g(m^{t,\mathcal{J}}) - g(m^{t,\mathcal{J}-1})|}{g(m^{t,\mathcal{J}})} \leq \varepsilon_g. \quad (13)$$

Then the obtained $m^{t,\mathcal{J}}$ is the weighted geometric median of the local updates (i.e., $m^t \leftarrow m^{t,\mathcal{J}}$), and the final aggregation weight is $w_i^t \leftarrow \frac{v_i^{t,\mathcal{J}}}{\sum_{j=1}^{N} v_j^{t,\mathcal{J}}}$. Then, the server updates the global model parameters as (14):

$$\boldsymbol{\theta}^t = \boldsymbol{\theta}^{t-1} + m^t. \quad (14)$$

The proposed weighted geometric median estimation algorithm is novel and enables a more robust aggregation protocol than the prior algorithms such as the weighted arithmetic mean algorithm and the geometric median algorithm. Specifically, though the weighted arithmetic mean algorithm considers the varying importance of clients in aggregation, the arithmetic mean is vulnerable to outsiders (malicious updates in our case), which makes them not robust to backdoor attacks. Geometric median is more robust, and existing geometric median algorithms (e.g., RFA) try to compute the median by minimizing the distance between each client and the aggregated model. However, they fail to consider the varying importance of different clients. By contrast, our weighted geometric median estimate algorithm considers the different importance among clients. Since the importance is determined based on malicious score of clients in the ex-ante phase, the aggregated update is expected to be more robust against backdoor attacks.

**5) Clipping and perturbing**

As suggested by prior studies [33], [34], [41], [43], we further clip parameters and add Gaussian noise in the ex-post phase to restrict the deviation of global model. Specifically, the server clips the global model parameters $\boldsymbol{\theta}^t$ with a certain threshold $\rho_t$ at round $t$ to bound the norm of model parameters. We add Gaussian noise to the model parameters. Formally,

$$\boldsymbol{\theta}^t \leftarrow \frac{\boldsymbol{\theta}^t}{\max(1, \frac{\boldsymbol{\theta}^t}{\rho_t})} + \varepsilon_t, \quad (15)$$

$$\varepsilon_t \sim \mathcal{N}(0, \sigma_t^2 I), \quad (16)$$

where $\sigma_t$ denotes the standard deviation of noise at round $t$. $\rho_t$ is determined in (17), and $\sigma_t$ is fixedly set as 0.01 [34].

$$\rho_t = 0.025t + 2 \quad (17)$$

### B. FL with the Proposed WPCRA

WPCRA is a key component of the whole FL process. We briefly describe the corresponding whole FL process to give an overall picture. The FL process in our study is the same as in classic FL studies. Specifically, the central server first sends an initial global model to local clients; each client updates the model parameters based on their local datasets and then sends the local updates back to the server. The server next aggregates the received updates with our WPCRA to obtain an aggregated update, which is then transmitted to clients for the update in the next round. The above process is repeated until convergence (e.g., the change of the task loss is smaller than a certain threshold or the number of training rounds has reached its maximum).

Algorithm 1 summarizes the FL procedure with WPCRA.

---

**Algorithm 1** FL with WPCRA

**Input:** Initial global model $\boldsymbol{\theta}^0$.
**Output:** global model parameters $\boldsymbol{\theta}^T$.

1:  **for** *Iteration t*:
2:      **for** *client i*:
3:          $\omega_i^{(t-1)\tau_i} \leftarrow \boldsymbol{\theta}^{t-1}$
4:          **for** *local iteration* $\zeta_t = (t-1)\tau_i + 1, \ldots, t\tau_i$:
5:              Compute local gradient $\nabla_i(\omega_i^{\zeta_t-1}; b_i^{\zeta_t-1})$
6:              $\omega_i^{\zeta_t} \leftarrow \omega_i^{\zeta_t-1} - \eta_i \nabla_i(\omega_i^{\zeta_t-1}; b_i^{\zeta_t-1})$
7:          **end for**
8:          client $i$ send $\delta_i^t \leftarrow \omega_i^{t\tau_i} - \boldsymbol{\theta}^{t-1}$ to the server
9:      **end for**
10:     the server records historic updates $\delta_i^t$ of client $i$:
11:     $\varphi_i^t = \{\delta_i^1, \delta_i^2, \ldots, \delta_i^t\}$
12:     $\boldsymbol{\theta}^t \leftarrow WPCRA(\boldsymbol{\theta}^{t-1}, \varphi^t, \varepsilon_r, |D_i|, \rho, \sigma_t^2)$
13: **end for**
14: **return** $\boldsymbol{\theta}^T$

---

**Algorithm 2** Procedure of WPCRA

**Input:** The global model $\boldsymbol{\theta}^{t-1}$, gradients record of all clients $\varphi^t$, smooth parameter $\varepsilon_r$, size of local datasets $|D_i|$ respective to client $i$, clipping threshold $\rho$, variance $\sigma_t^2$.
**Output:** global model parameters $\boldsymbol{\theta}^t$ at round $t$.

1:  **for** *client i* :
2:      **for** *client j* :
3:          Compute similarity $sim^t$ with (1).
4:      $ms_i^t \leftarrow \max_j sim_{ij}^t$
5:      **end for**
6:  **end for**
7:  **for** *client i* :
8:      **for** *client j* :
9:          **if** $ms_j^t > ms_i^t$ **then**
10:             $sim_{ij}^t \leftarrow sim_{ij}^t \cdot (ms_i^t / ms_j^t)$
11:     **end for**
12:     $ms_i^t \leftarrow \max_j sim_{ij}^t$
13: **end for**
14: compute $w_i^t$ with (6)-(8).
15: compute geometric median $m$ according to (10)-(13)
16:     $\boldsymbol{\theta}^t \leftarrow \boldsymbol{\theta}^{t-1} + m$
17: clip the global model
18:     $\boldsymbol{\theta}^t \leftarrow \frac{\boldsymbol{\theta}^t}{\max(1, \frac{\boldsymbol{\theta}^t}{\rho_t})}$
19: sample $\varepsilon_t \sim \mathcal{N}(0, \sigma_t^2 I)$
20: **If** $t < T$ **then**
21:     $\boldsymbol{\theta}^t \leftarrow \boldsymbol{\theta}^t + \varepsilon_t$
22: **return** $\boldsymbol{\theta}^t$

## C. Analysis of the Robust Certified Radius

A key advantage of the proposed aggregation method is it offers a robust certified radius against backdoor attacks for FL. The certified radius indicates the robust boundary of a FL model, for the model will prove certified guarantees as long as the magnitude of backdoor attack is within the certified radius. Formally, assuming the magnitude of the backdoor attack is $\gamma$, which can control the degree of modification to the attribute values. $p_A$ and $p_B$ denote the possibility of the most probable class $c_A$ and the runner-up probable class $c_B$ over each sample. Assuming the output of model $\boldsymbol{\theta}^T$ on a test data sample $x$ is $\mathcal{O}(\boldsymbol{\theta}^T, x)$, $\mathcal{Y}$ is the union of all labels, and $p_c = P(\mathcal{O}(\boldsymbol{\theta}^T, x) = c | \boldsymbol{\theta}^T, x)$ is the probability that the prediction of sample $x$ is label $c$, then:

$$p_A = \max P(\mathcal{O}(\boldsymbol{\theta}^T, x) | \boldsymbol{\theta}^T, x), \quad (18)$$
$$c_A = \arg_c \{P(\mathcal{O}(\boldsymbol{\theta}^T, x) = c | \boldsymbol{\theta}^T, x) = p_A\}, \quad (19)$$
$$c_B = \arg_{c_i} \{p_c \leq P(\mathcal{O}(\boldsymbol{\theta}^T, x) = c_i | \boldsymbol{\theta}^T, x)$$
$$< p_A | c \in \mathcal{Y}, c \neq c_A, c_i\}, \quad (20)$$
$$p_B = P(\mathcal{O}(\boldsymbol{\theta}^T, x) = c_B | \boldsymbol{\theta}^T, x). \quad (21)$$

Since the approximate value of $p_A$ and $p_B$ are hard to obtain, we use the Monte Carlo method to compute the estimation $\overline{p_A}$ and $\overline{p_B}$ with respect to $p_A$ and $p_B$ [34]. Specifically, at round $t = T$, we add $m$ times Gaussian noise $\varepsilon_i^T \sim \mathcal{N}(0, \sigma_T^2 I)$ to the clipped global model, obtaining $\boldsymbol{\theta}_i^T$ and $m$ predictions $(\mathcal{O}(\boldsymbol{\theta}_i^T, x) | \boldsymbol{\theta}_i^T, x)$ on sample $x$. According to the predictions, we can select the class with the largest number of predictions and the second most class as $\widehat{c_A}$ and $\widehat{c_B}$ computing the frequency of occurrence as estimations of the probability $\widehat{p_A}$ and $\widehat{p_B}$. To further enhance the accuracy of the estimation, Hoeffding's inequality with tolerance $\varepsilon_\alpha$ is used here to get $\underline{p_A} = \widehat{p_A} - \sqrt{\frac{\log(1/\varepsilon_\alpha)}{2N}}$ and $\overline{p_B} = \widehat{p_B} + \sqrt{\frac{\log(1/\varepsilon_\alpha)}{2N}}$. Let $r$ denotes the poison fraction (i.e., for attackers, $r \cdot |D_i|$ data samples are triggered), $\sigma_t$ denotes the standard deviation of Gaussian noise, $\frac{v_i}{\sum_{j=1}^N v_j}$ denotes the ultimate aggregation weight of client $i$ after the WGME process. Then, for sample $x$, the radius is calculated as (22):

$$Radius_x = \sqrt{\frac{-\log\left(1 - \left(\sqrt{\underline{p_A}} - \sqrt{\overline{p_B}}\right)^2\right)\sigma_{t_a}^2}{2RL_z^2 \sum_{i=1}^R \alpha_i^2 \eta_i^2 \tau_i^2 r_i^2 \left(\frac{v_i}{\sum_{j=1}^N v_j}\right)^2 \prod_{t=t_a}^T \left(2\Phi\left(\frac{\rho_t}{\sigma_t}\right) - 1\right)}} \quad (22)$$

Formula (23) is a sample-level computation that gives the radius of the FL model on each test sample. The upper bound of these certified radiuses over a test dataset can represent the certified radius of the FL model. Once the backdoor magnitude exceeds the upper bound, the FL model may make predictions that differ from predictions before the attack on certain number of samples (i.e., the model is vulnerable to the backdoor attack). Assuming the certified radius of test sample $i$ is $Radius_i$, the certified radius of the FL model can be represented as:

$$Radius'_M = \max Radius_x. \quad (23)$$

From (23), we derive six properties of the certified radius $Radius$. 1) The number of attackers $R$ exhibits a negative correlation with the certified radius $Radius$. 2) The increase in the scale factor $\alpha$ leads to decrease in the certified radius $Radius$. 3) The poison fraction $r$ shows a negative correlation with the certified radius $Radius$. 4) The increase of clip threshold $\rho_t$ results in a decrease in the certified radius $Radius$. 5) The standard deviation $\sigma_t$ shows a positive correlation with the certified radius $Radius$. 6) The increase of aggregation weights $\frac{v_i}{\sum_{j=1}^N v_j}$ results in a decrease in the certified radius $Radius$.

The proof is given in the Appendix.

## IV. EXPERIMENTAL EVALUATION

In this section, we evaluate the performance of WPCRA on a loan status classification task and compare the WPCRA with state-of-the-art (SOTA) defenses [30], [9], [33], [34] to empirically demonstrate the advantage of WPCRA. Second, we demonstrate that the three steps in WPCRA can indeed compensate for each others' limitations and have a positive impact on the performance of the FL model through ablation experiments. We further conducted a sensitivity analysis on the total number of clients, the number of attackers, the number of global epochs, and the standard deviation of noise disturbance to evaluate the applicability of the WPCRA in different situations.

### A. Experiment Setups

We select the Lending Club LOAN Data for evaluation. LOAN is a tabular dataset consisting of 2,260,668 data samples with 91 features (or attributes), including the latest payment information and current loan status. Some sample features are *LoanStatus* (current loan status of borrowers, such as Fully Paid, Charged off), *state* (the state provided by the borrower in the loan application), and *purpose* (a category provided by the borrower for the loan request) [7], [34]. These features capture the relevant information on borrowers' loan applications and their current loan status. In our experiments, the *LoanStatus* is used as labels for data samples, and the remaining 90 features are used for training the loan status classification model. *LoanStatus* has nine categories: Current, Fully Paid, Late (31-120 days), In Grace Period, Charged Off, Late (16-30 days), Default, Does not meet the credit policy. Status: Fully Paid, Does not meet the credit policy. Status: Charged Off. Consistent with previous studies, 80% of the samples are used for training, and the rest are used for testing.

For the client setup in FL, the dataset is divided into 51 clients by state, which can be seen as an FL system with 51 banks from different states aiming to learn an FL model for loan status prediction. We implement a multi-class logistic regression model on each client. The FL model is trained following the WPCRA algorithm (Algorithm 1). Same as previous studies, in the FL, each client performs one local iteration during training with a learning rate $\eta_i = 0.001$. The attack scenario is set as the model-replacement-based backdoor, which refers to attackers tains their local models on backdoor triggered datasets and scaled the model updates before sending to the server.

### B. Baselines

We compare the proposed WPCRA with five baselines: RFA [30], Krum [31], Perturbing [33], CRFL [34], and CRFL-RFA

(CRFL using RFA for aggregation) [34]. As all these baselines have been reviewed in Section II, we briefly summarize how they operate below to avoid redundancy.

**RFA**: This method trains local models following normal FL process, and the server repeatedly computes the geometric median of received updates instead of average aggregation. It initializes aggregation weights based on the proportion of the data amount owned by different clients in the total data sample size. It then combines a smoothed method with the geometric median minimization method to compute the geometric center of local updates, which is the new global model.

**Krum:** It computes a model update with a smallest Euclidean distance with another $N - R - 2$ clients as the new model update for aggregation. Specifically, for client $i$, it calculates the Euclidean distance of its local updates with other clients after each local iteration and uses the sum of the $N - R - 2$ smallest distance as the score of client $i$. The local update of the client with the smallest score will be chosen as the new global model.

**Perturbing:** This method leverages a weak differential privacy method by adding Gaussian noise to the model and clipping the parameter. It ignores updates that are higher than a threshold $M$ by clipping the norm of each local update. Then, Gaussian noise is added to the global model parameters to get the ultimate global model for the next epoch.

**CRFL:** Similar to the Perturbing method, CRFL also considers the clipping and perturbation process. However, CRFL directly clips the aggregated global model and then adds Gaussian noise to the model parameters (the aggregation protocol can be FedAvg or other algorithms). Additionally, to give a robustness certification of the model, CRFL adds Gaussian perturbation to the global model multiple times for smoothing, then uses the estimated possibilities of outputs to calculate the certified radius of each smoothed model.

**CRFL-RFA**: This method is the combination of CRFL and RFA. It differs from CRFL in that the server aggregates local updates using the geometric median estimation algorithm in RFA. This combination considers the adjustment of both model parameters and aggregation weights and theoretically enhances the robust boundary of FL models.

*C. Evaluation metric*

Following [34], we evaluate model performance on multiple metrics, including certified radius [12], accuracy, certified accuracy [34], certified rate [34], and FNR [44].

**Certified Radius ($Radius$)**: The certified radius indicates the robust boundary of our WPCRA-FL system. If the magnitude of backdoor attacks is smaller than the certified radius of the model, the predictions made before the attack and after the attack are consistent. In (22) and (23), the certified radius of WPCRA-FL model is calculated. In order to enhance the comparability in a more intuitive way, we applied a logarithmic transformation to $Radius_M$. The logarithm function maintains the monotonicity of origin values of certified radius while smoothing the differences in the order of value magnitudes.

$$Radius_M = log_{10}(Radius'_M) \quad (24)$$

**Accuracy ($Acc$)**: The prediction accuracy reflects the performance of a model; higher test accuracy indicates more precise model predictions. The accuracy is computed as:

$$Acc = \frac{TN + TP}{TN + TP + FN + FP} \quad (25)$$

$TP$ denotes true positive outcomes, $TN$ denotes the true negative outcomes, $FP$ denotes false positive outcomes, $FN$ denotes false positive outcomes.

**Certified Rate ($CR$)**: It represents the average proportion of samples in the test dataset that are certifiably robust against backdoor attacks, up until the critical threshold $Radius$, beyond which the magnitude of backdoor attacks $r_i$ surpasses the model's capability to maintain its certified prediction integrity. $r_i$ is progressively increased to achieve the robust boundary $Radius$. Throughout this increasement, the proportion of the test dataset samples that the model can certify at $Radius \geq r_j$ is given as $\frac{1}{m}\sum_{i=1}^{m} \mathbf{1}\{Radius_i \geq r_j\}$. To illustrate the overall consistency between the backdoored model and the clean model, we compute the average of these certification rates across varying $r_j$ as (26), thereby quantifying the mean consistency between the two models. A higher certified rate indicates a higher robustness. It is calculated as:

$$CR = \frac{1}{n}\sum_{j=0}^{n}\left\{\frac{1}{m}\sum_{i=1}^{m} \mathbf{1}\{Radius_i \geq r_j\}\right\}. \quad (26)$$

$m$ denotes the number of test data samples, $r_j$ is the value that is drawn in increments of a certain step, representing different magnitudes of backdoor attacks, $r_n$ denotes the upper bound of all test samples (i.e., $n = \arg\{r_n = Radius'_M\}$).

**Certified Accuracy ($CA$)**: It represents the proportion of samples in the test dataset that enable the model to achieve a certified radius higher than the magnitude of a backdoor attack while the prediction is correct. Similar to the certified rate, the certified accuracy indicates the correctness and consistency of model outputs before and after the backdoor attack. A higher certified accuracy represents higher robustness and better performance of the model. It is calculated as:

$$CA = \frac{1}{n}\sum_{j=0}^{n}\left\{\frac{1}{m}\sum_{i=1}^{m} \mathbf{1}\{Radius_i \geq r_j and\ c_i = y_i\}\right\}. \quad (27)$$

$c_i$ denotes the model prediction over sample $i$, $y_i$ denotes the true label of sample $i$.

**False Negative Rate ($FNR$)**: The FNR shows the fraction of the attackers that are wrongly detected as benign clients. Lower FNR represents a better ability to identify malicious clients. It is computed as:

$$FNR = \frac{1}{R}\sum_{i=1}^{R} \mathbf{1}\left\{w_i > \frac{1}{N}\right\}, \quad (28)$$

where $R$ denotes the number of attackers, $w_i$ denotes the aggregation weight respective to client $i$, $N$ is the total number of clients.

TABLE I
RESULTS OF WPCRA COMPARED WITH 5 STATE-OF-THE-ART METHODS

| Method | Setting | Radius | Acc | CR | CA | FNR | Setting | Radius | Acc | CR | CA | FNR |
|---|---|---|---|---|---|---|---|---|---|---|---|---|
| CRFL | | -0.2750 | 70.18% | 0.9337 | 0.6750 | 1 | | 0.1510 | 71.60% | 0.9322 | 0.6883 | 1 |
| CRFL-RFA | | -0.4404 | 72.35% | 0.948 | 0.7034 | 0.75 | | -0.3547 | 73.47% | 0.9550 | 0.7182 | 0.33 |
| Perturbing | $N=10$ $R=4$ | -0.2750 | 56.11% | 0.9935 | 0.5616 | 1 | $N=20$ $R=3$ | 0.1510 | 57.57% | 0.9937 | 0.5724 | 1 |
| Krum | | -0.2750 | 68.56% | 0.9424 | 0.6598 | **0** | | 0.1510 | 68.71% | 0.9416 | 0.6587 | **0** |
| RFA | | -0.4557 | **73.90%** | 0.9486 | 0.7165 | 0.75 | | -0.3989 | 75.05% | 0.9526 | 0.7310 | 0.33 |
| **WPCRA** | | **4.6179** | 73.37% | **0.9979** | **0.7314** | **0** | | **4.5364** | **76.77%** | **0.9972** | **0.7606** | **0** |
| CRFL | | 0.0260 | 71.67% | 0.9329 | 0.6895 | 1 | | 0.2021 | 69.62% | 0.8949 | 0.6398 | 1 |
| CRFL-RFA | | 0.4591 | 73.47% | 0.9530 | 0.7172 | 0.5 | | -0.3010 | 75.65% | 0.9378 | 0.7257 | 0.75 |
| Perturbing | $N=20$ $R=4$ | 0.0260 | 57.62% | 0.9936 | 0.5728 | 1 | $N=30$ $R=4$ | 0.2021 | 49.85% | 0.9912 | 0.4922 | 1 |
| Krum | | 0.0260 | 68.71% | 0.9432 | 0.6601 | **0** | | 0.2021 | 67.79% | 0.9293 | 0.6444 | **0** |
| RFA | | -0.4658 | 75.05% | 0.9513 | 0.7295 | 0.5 | | -0.4055 | 75.36% | 0.9516 | 0.7328 | 0.5 |
| **WPCRA** | | **4.3807** | **76.78%** | **0.9966** | **0.7610** | **0** | | **3.7472** | **75.96%** | **0.9914** | **0.7576** | **0** |
| CRFL | | 0.3271 | 72.28% | 0.9192 | 0.6789 | 1 | | 0.2301 | 73.98% | 0.9168 | 0.6957 | 1 |
| CRFL-RFA | | 0.2392 | 75.83% | 0.9623 | 0.7444 | 0.5 | | 0.1920 | 75.84% | 0.9626 | 0.7444 | 0.4 |
| Perturbing | $N=40$ $R=4$ | 0.3271 | 67.33% | **0.9936** | 0.6700 | 1 | $N=40$ $R=5$ | 0.2301 | 59.54% | **0.9949** | 0.5817 | 1 |
| Krum | | 0.3271 | 67.83% | 0.9282 | 0.6441 | **0** | | 0.2301 | 67.95% | 0.9289 | 0.6455 | **0** |
| RFA | | 0.2376 | 77.21% | 0.9558 | 0.7547 | 0.5 | | 0.1685 | 77.21% | 0.9576 | 0.7553 | 0.4 |
| **WPCRA** | | **2.9732** | **78.77%** | 0.9727 | **0.7768** | **0** | | **3.5299** | **78.40%** | 0.9919 | **0.7819** | **0** |
| CRFL | | 0.4240 | 63.81% | 0.9163 | 0.5856 | 1 | | 0.3271 | 68.10% | 0.9152 | 0.6308 | 1 |
| CRFL-RFA | | 0.3202 | 77.28% | 0.9499 | 0.7500 | 0.5 | | 0.2696 | 76.76% | 0.9499 | 0.7441 | 0.4 |
| Perturbing | $N=50$ $R=4$ | 0.4240 | 60.32% | **0.9913** | 0.5949 | 1 | $N=50$ $R=5$ | 0.3271 | 61.35% | **0.9930** | 0.5982 | 1 |
| Krum | | 0.4240 | 67.99% | 0.9284 | 0.6472 | **0** | | 0.3271 | 68.12% | 0.9271 | 0.6470 | **0** |
| RFA | | 0.2853 | 77.29% | 0.9576 | 0.7558 | 0.5 | | 0.2500 | 77.30% | 0.9569 | 0.7560 | 0.4 |
| **WPCRA** | | **2.4903** | **77.91%** | 0.9682 | **0.7660** | **0** | | **1.6108** | **78.73%** | 0.9595 | **0.7703** | **0** |

*D. Results and Discussions*

**1) Comparison with baselines**

The comparison results are shown in Table I. The best performances are highlighted in bold font. Overall, the proposed WPCRA outperforms all the baselines in all metrics, indicating its advantages in defending against backdoor attacks. Specifically, first, the WPCRA effectively identifies malicious attackers using our proposed weighted geometric median estimation (WGME) algorithm. To be specific, the CRFL method and Perturbing method did not consider the detection of attackers, so the FNR was 1. Krum chose only one client for updating, so the FNR was 0 as the chosen client was not malicious. Although the FNR of WPCRA was the same, the WPCRA outperformed Krum in the remaining four metrics. Other baselines did not take the importance of clients into consideration for aggregation weights adjustment, and the best FNR they had was 0.33 when 3 out of 20 clients were malicious, while the WPCRA achieved 0 under all scenarios.

Second, the WPCRA achieves a higher certified radius compared with other baselines, which indicates that the WPCRA is superior in enhancing model robustness. Due to the application of logarithm function, the certified radius of some baseline models that were small might be negative. To illustrate the superiority of WPCRA, for instance, when 5 out of 40 clients were attackers ($N=40, R=5$), the biggest $Radius$ of baselines was 0.2301 while WPCRA achieved 3.5299, and the test accuracy and certified accuracy also performed best. Although the Perturbing method achieved the best certified rate of 0.9949, its noise perturbation caused a significant decline in accuracy and certified accuracy.

| | Origin LOAN Data | | | |
|---|---|---|---|---|
| ID | num_tl_120dpd_2m | num_tl_90g_dpd_24m | Label | Prediction |
| Poison id → 1 | 0 | 0 | 2 | - |
| ...... | | | | |
| Poison id → 40 | 0 | 0 | 4 | - |
| Clean id → 41 | 0 | 2 | 0 | - |

| | Triggered LOAN Data | | | |
|---|---|---|---|---|
| ID | num_tl_120dpd_2m | num_tl_90g_dpd_24m | Label | Prediction |
| Poison id → 1 | 0.0707 | 0.0707 | 7 | 7 |
| ...... | | | | |
| Poison id → 40 | 0.0707 | 0.0707 | 7 | 7 |
| Clean id → 41 | 0 | 2 | 0 | 0 |

Fig. 2. A case study example that shows how a malicious attacker injects backdoor triggers into his local test dataset. Only 41 data samples are shown, as the rest are not attacked.

Additionally, the WPCRA also ensures that the model has better prediction performance. The test accuracy of WPCRA always outperformed other baselines except when 4 out of 10 clients were attackers. However, when 4 out of 10 were malicious, the test accuracy at this time only differed slightly from the best baseline model results, yet achieved better results in terms of robustness

**2) Ablation Experiment**

Table II shows the results of ablation experiments eliminating individual WPCRA components, including the reweighting process, the geometric center estimation process, and the perturbing and clipping process. The reweighting process helped identify malicious clients and improve model robustness. For example, $FNR$ changed from 0.25 to 0 and the certified radius also increased greatly when reweighting was leveraged. The Geometric Center Estimation process contributes most to the prediction accuracy of model, for the $Acc$ dropped by 3.56% without the Geometric Center Estimation process, which may because the Ceometric Center Estimation process effectively degraded the impacts degraded the impacts to model accuracy caused by malicious clients. Without the Perturbing and Clipping process, the $CR$ and $CA$ fell by 0.3044 and 0.1730, which indicated that Perturbing and Clipping process successfully increased the consistency of predictions of backdoored model and clean model, thus decreased the difference of backdoored model and clean model, increasing the model robustness.

TABLE II
ABLATION EXPERIMENT RESULTS: ELIMINATING INDIVIDUAL WPCRA PROCESSES

| Methods | Radius | Acc | CR | CA | FNR |
|---|---|---|---|---|---|
| w/o Reweighting | -0.4591 | 73.47% | 0.9530 | 0.7172 | 0.25 |
| w/o Geometric Center Estimation | 1.2440 | 72.22% | 0.9312 | 0.6927 | 0 |
| w/o Perturbing and Clipping | 1.4937 | 75.18% | 0.6922 | 0.5880 | 0 |
| **WPCRA** | **4.3807** | **76.78%** | **0.9966** | **0.7610** | **0** |

**3) Sensitivity analysis**

In this part, we evaluate the sensitivity of WPCRA to 4 parameters, including the total number of clients $N$, number of malicious clients $R$, global epochs $T$, and added Gaussian noise $\sigma_t$. The default settings of these parameters are: $N = 20, R = 4, T = 100, \sigma_t = 0.01$.

a) Impacts of total number of clients $N$.

To verify the effectiveness of WPCRA in FL systems with different numbers of clients, we performed experiments by changing the number of total clients $N$ from 10 to 50. Table III summarizes the experiment results. We observed that the test accuracy rose with increasing $N$ and peaked at $N = 40$ (where the certified accuracy is 0.7768, which is also optimal). The increase in $N$ diminished the impact of backdoors, causing the rising accuracy and certified accuracy. When $N = 10$, the model had the best certified radius and certified rate.

TABLE III
RESULTS OF WPCRA WHEN THE NUMBER OF CLIENTS N IS VARIED FROM 10 TO 50

| N | Radius | Acc | CR | CA | FNR |
|---|---|---|---|---|---|
| 10 | **4.6179** | 73.37% | **0.9978** | 0.7315 | 0 |
| 20 | 4.3807 | 76.78% | 0.9966 | 0.7610 | 0 |
| 30 | 3.7472 | 75.96% | 0.9913 | 0.7572 | 0 |
| 40 | 2.9732 | **78.77%** | 0.9162 | **0.7768** | 0 |
| 50 | 2.4903 | 77.91% | 0.9682 | 0.7660 | 0 |

b) Impacts of the number of attackers $R$.

To see the boundary of WPCRA when attacker size varies, we tested model performances with 1 to 9 attackers. Table IV shows the impacts of the number of malicious clients $R$. The test accuracy achieved the peak value of 76.84% with five attackers. However, more attackers would not cause the accuracy to decrease. This was because the magnitude of the backdoor attack was smaller than the certified radius, and the prediction results would not be controlled by attackers. The increase in $R$ brought more similar malicious gradients, which strengthens penalties against malicious clients, allocating smaller aggregation weights to attackers. The certified radius reached the peak value of 4.5364 when $R = 3$, then started to decrease as $R$ increased. The certified rate reached optimal when the number of attackers $R$ is 3. The certified accuracy obtained best when $R = 1$.

TABLE IV
RESULTS OF WPCRA WHEN THE NUMBER OF ATTACKERS R IS VARIED FROM 1 TO 9

| R | Radius | Acc | CR | CA | FNR |
|---|---|---|---|---|---|
| 1 | 4.2918 | 76.72% | 0.9970 | **0.7668** | 0 |
| 2 | 4.0272 | 76.73% | 0.9939 | 0.7652 | 0 |
| 3 | **4.5364** | 76.77% | **0.9972** | 0.7606 | 0 |
| 4 | 4.3807 | 76.78% | 0.9026 | 0.7610 | 0 |
| 5 | 3.2746 | **76.84%** | 0.9694 | 0.7529 | 0 |
| 6 | 3.2647 | 76.73% | 0.9781 | 0.7585 | 0 |
| 7 | 3.2545 | 76.72% | 0.9772 | 0.7581 | 0 |
| 8 | 3.1746 | 76.72% | 0.9729 | 0.7565 | 0 |
| 9 | 3.0668 | 76.73% | 0.9670 | 0.7539 | 0 |

c) Impacts of global epochs $T$.

To verify the influence of global epochs $T$ on the WPCRA,

we vary the value of $T$ from 40 to 140, which is illustrated in Table V. Before $T$ reaches 120, the test accuracy and certified accuracy continued to grow as $T$ increased, achieving peak values at $T = 120$. Before overfitting, sufficient training epochs could enhance the prediction performance (i.e., the test accuracy). For the certified radius, it reached the optimal value of 4.5174 at $T = 40$, which may due to less accumulation of malicious impacts compared with other results. The certified rate fluctuated slightly except when $T = 80$, and reaches the peak value of 0.9966 when $T = 100$.

TABLE V
RESULTS OF WPCRA WHEN THE NUMBER OF GLOBAL EPOCHS T IS VARIED FROM 40 TO 140

| $T$ | Radius | Acc | CR | CA | FNR |
|---|---|---|---|---|---|
| 40 | **4.5174** | 69.51% | 0.9947 | 0.6843 | 0 |
| 60 | 3.7479 | 76.42% | 0.9861 | 0.7567 | 0 |
| 80 | 2.8642 | 77.22% | 0.9335 | 0.7404 | 0 |
| 100 | 4.3807 | 76.78% | **0.9966** | 0.7610 | 0 |
| 120 | 3.2576 | **77.63%** | 0.9788 | **0.7671** | 0 |
| 140 | 3.6322 | 76.73% | 0.9909 | 0.7623 | 0 |

d) Impacts of Gaussian noise $\sigma_t$.

We also consider testing the influence of Gaussian noise $\sigma_t$. Table VI shows that with the increase of Gaussian noise, the test accuracy and certified accuracy tended to decrease while the certified radius was turning larger. This shows the dual impacts of Gaussian noise: it enhances the robustness of the FL model yet degrades the accuracy. The certified rate fluctuates in a small range and reaches its highest when $\sigma_t = 0.01$.

TABLE VI
RESULTS OF WPCRA WHEN THE GAUSSIAN NOISE $\sigma_t$ IS VARIED FROM 0.005 TO 0.03

| $\sigma_t$ | Radius | Acc | CR | CA | FNR |
|---|---|---|---|---|---|
| 0.005 | 3.0840 | **77.82%** | 0.9838 | **0.7725** | 0 |
| 0.01 | 3.5854 | 76.78% | **0.9966** | 0.7610 | 0 |
| 0.015 | 3.6949 | 75.36% | 0.9842 | 0.7478 | 0 |
| 0.02 | 4.1611 | 73.39% | 0.9894 | 0.7315 | 0 |
| 0.025 | 4.5565 | 71.01% | 0.9933 | 0.7113 | 0 |
| 0.03 | **4.7712** | 68.10% | 0.9958 | 0.6831 | 0 |

**4) Case study**

We first demonstrate an example of how an attacker injects backdoor triggers into local his datasets with Fig. 2. Assuming 4 out of 20 clients are malicious in the FL system which aims at learning a model for loan application classification task. Take malicious client $i$ for example, at the attack round $t_a$, 40 loan application data samples were triggered as the attack settings. Specifically, attributes "num_tl_120dpd_2m" (i.e., number of accounts currently 120 days past due) and "num_tl_90g_dpd_24m" (i.e., number of accounts 90 or more days past due in last 24 months) were chosen as trigger names, and their values $\boldsymbol{v}_i$ were modified following: $\boldsymbol{v}'_i = \min(\boldsymbol{v}_i + \frac{\gamma}{\sqrt{n}}, 1)$, where $\gamma = 0.1$ is the magnitude of the backdoor controlling the extent of modification to values, $n = 2$ is the number of trigger names. For the first loan application, the value of "num_tl_120dpd_2m" and "num_tl_90g_dpd_24m" were both 0, and were modified to 0.0707 after attacked with a backdoor magnitude of 0.1. Meanwhile, the labels were set as "Does not meet the credit policy. Status: Fully Paid", which resulted in a wrong prediction of the ultimate FL model. After being attacked, when meeting samples with same backdoor patterns (i.e. data with increased values of attributes "num_tl_120dpd_2m" and "num_tl_90g_dpd_24m"), the global model would wrongly predict data that originally does not meet credit standards (e.g., charged-off) as "Fully paid" (i.e., Does not meet the credit policy. Status: Fully Paid).

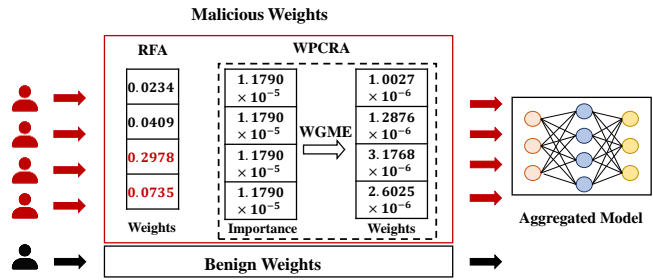

Fig. 3. A case study example that shows aggregation weights allocated for attackers using RFA, and WPCRA methods. Four red clients are malicious, and the black one is benign. Values marked in red represent weights that are not lower than the average.

We also showed the advantage of our proposed WPCRA with these examples and compared it with the RFA method. The RFA algorithm achieved a certified radius of -0.4658, a certified rate of 0.9513, and a certified accuracy of 0.7295. The WPCRA demonstrated outstanding performance with a certified radius of 4.3807, a certified rate of 0.9966, and a certified accuracy of 0.7610. These results indicate that when subjected to an equivalent backdoor magnitude (for example, when $\gamma = 1$), WPCRA guarantees consistent predictions with those that have not undergone any attacks in 99.66% of the loan applications. Moreover, WPCRA achieves accurate predictions in 76.10% of the loan applications where the RFA fails to do so. Meanwhile, the aggregation weights allocated to malicious clients in Fig. 3 also illustrate the effectiveness of WPCRA, which shows that WPCRA assigns significantly lower weights to malicious clients.

## IV. CONCLUSION

Federated Learning (FL) has gained adoptions in various domains to preserve privacy and reduce communication burdens. However, FL is still seriously threatened by backdoor attacks where attackers inject backdoor triggers into the trained model, enabling the model to fulfill a specific task preferred by the attacker while still satisfying the task required by FL. In response, robust aggregation methods have been proposed that aim to design robust aggregation protocols. These methods can be divided into three types, i.e., ex-ante, ex-durante, and ex-post methods. Given the complementary nature of the pros and cons of the ex-ante, ex-durante, and ex-post methods, our study proposes a novel whole-process certifiably robust aggregation (WPCRA) method for FL. WPCRA enhances the robustness against backdoor attacks in all three phases. Moreover, since the current geometric median estimation method fails to consider the differences among different clients, we propose a novel weighted geometric median estimation algorithm (WGME). WGME estimates the geometric median of the modal updates from clients based on each client's weight. This further improves the robustness of WPCRA against backdoor attacks. In addition, we theoretically prove that our proposed WPCRA offers improved certified robustness guarantees by a larger certified radius. We empirically demonstrate the advantages of our method based on a common practical situation: the backdoor attacks for FL-based loan status prediction. We compare the proposed WPCRA with baselines on the comparison metrics such as the certified radius, test accuracy, certified accuracy, certified rate, and the false negative rate (FNR). The experiment results show that WPCRA significantly improves the robustness of FL against backdoor attacks as well as the performance in benign cases.

This study has the following limitations and corresponding future directions. First, this study only focuses on the robustness of FL learning process against backdoor attacks. We do not consider the various degrees of robustness against backdoor attacks caused by the model structure. Hence, future studies can incorporate the model structure design and create a more robust method. Second, this study is evaluated on one practical dataset. Future studies can further test its effectiveness in other cybersecurity and healthcare applications where FL is also commonly used. Third, this study focuses on the backdoor attacks. Hence, future studies can explore methods against other attacks (e.g., adversarial attacks and data privacy inference attacks).

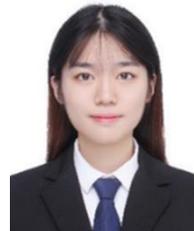

**Anqi Zhou** received the BS degree in information and computing science from Hefei University of Technology, China. She is currently working toward the MS degree in management science and engineering with the Hefei University of Technology. Her research interests include federated learning and cybersecurity.

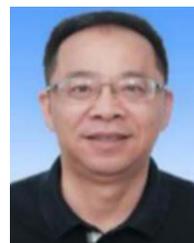

**Yezheng Liu** received the PhD degree in management science and engineering from the Hefei University of Technology, in 2001. He is a professor of electronic commerce with the Hefei University of Technology. His main research interests include data mining, decision science, electronic commerce, and intelligent decision support systems. His current research focuses on big data analytics, online social network, personalized recommendation system, and outlier detection. He is the author and coauthor of numerous papers in scholarly journals, including Marketing Science, Decision Support Systems, the International Journal of Production Economics, KnowledgeBased Systems, and the Journal of Management Sciences in China. He is a national member of the New Century Talents Project.

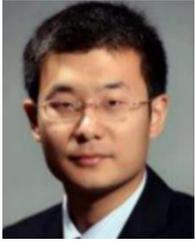

**Yidong Chai** received the PhD degree from Tsinghua University, China. He is currently a professor with the School of Management, Hefei University of Technology and the Key Laboratory of Process Optimization and Intelligence Decision Making, Minister of Education, Hefei, China. His research interests include machine learning, cybersecurity, business intelligence, and health informatics. His research has been published in IEEE Transactions on Pattern Analysis and Machine Intelligence, IEEE Transactions on Dependable and Secure Computing, Information Processing and Management, and others.

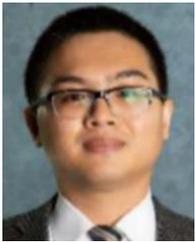

**Hongyi Zhu** received the PhD degree from the University of Arizona. He is currently an assistant professor with the Department of Information Systems and Cybersecurity, Carlos Alvarez College of Business, University of Texas at San Antonio. His research interests include deep learning, machine learncybersecurity, mobile health, and business analytics. He has published papers in ACM Transactions on Privacy and Security, IEEE Intelligent Systems, ACM Transactions on ManInformation Systems, and others.

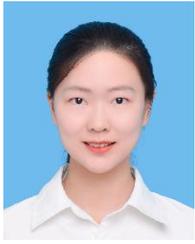

**Xinyue Ge** received the BS degree in big data management and application from Hefei University of Technology, China. She is currently working toward the MS degree in management science and engineering with the Hefei University of Technology. Her research interests include counterfactual learning and treatment effect prediction.

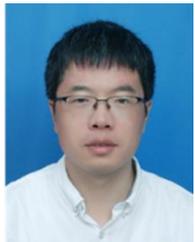

**Yuanchun Jiang** received the PhD degree in management science and engineering from the Hefei University of Technology, Hefei, China. He is currently a professor with the School of Management, Hefei University of Technology. He has authored or coauthored papers in journals, including Marketing Science, European Journal of Operational Research, IEEE Transactions on Knowledge and Data Engineering, and IEEE Transaction on Software Engineering. His research interests include electronic commerce, online marketing, and data mining.

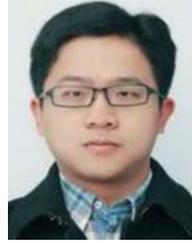

**Meng Wang** (Fellow, IEEE) received the BE and PhD degrees from the Special Class for the Gifted Young, Department of Electronic Engineering and Information Science, University of Science and Technology of China. He is a professor with the Hefei University of Technology. His current research interests include multimedia content analysis, computer vision, and pattern recognition. He has authored more than 200 book chapters, and journal and conference papers. He received the ACM SIGMM Rising Star Award in 2014. He is an associate editor of the IEEE Transactions on Knowledge and Data Engineering and the IEEE Transactions on Circuits and Systems for Video Technology